# Imaging the atomic-scale electronic states induced by a pair of hole dopants in $Ca_2CuO_2Cl_2$ Mott insulator

Haiwei Li[1,*], Shusen Ye[1,*], Jianfa Zhao[2], Changqing Jin[2], and Yayu Wang[1,3†]

[1]*State Key Laboratory of Low Dimensional Quantum Physics, Department of Physics, Tsinghua University, Beijing 100084, P. R. China*

[2]*Beijing National Laboratory for Condensed Matter Physics, Institute of Physics, Chinese Academy of Sciences, Beijing 100190, P. R. China*

[3]*Frontier Science Center for Quantum Information, Beijing 100084, P. R. China*

[*]These authors contributed equally to this work.

[†]Email: yayuwang@tsinghua.edu.cn

We use scanning tunneling microscopy to visualize the atomic-scale electronic states induced by a pair of hole dopants in $Ca_2CuO_2Cl_2$ parent Mott insulator of cuprates. We find that when the two dopants approach each other, the transfer of spectral weight from high energy Hubbard band to low energy in-gap state creates a broad peak and nearly V-shaped gap around the Fermi level. The peak position shows a sudden drop at distance around 4 $a_0$ and then remains almost constant. The in-gap states exhibit peculiar spatial distributions depending on the configuration of the two dopants relative to the underlying Cu lattice. These results shed important new lights on the evolution of low energy electronic states when a few holes are doped into parent cuprates.

One of the few consensuses regarding the cuprate high temperature superconductors is that the parent compound is a Mott insulator, in which the half-filled band derived from the Cu $3d_{x^2-y^2}$ orbital splits into the lower and upper Hubbard bands by strong onsite Coulomb repulsion [1]. Introducing charge carriers into the $CuO_2$ plane by chemical dopants leads to the emergence of exotic states of matter such as the pseudogap phase [2], density wave orders [3], and the superconducting phase. A crucial step in unveiling the mechanism of superconductivity is to elucidate the electronic structure evolution of the doped Mott insulator with increasing carrier density [4]. However, it turns out to be one of the most challenging tasks in condensed matter physics due to the existence of strong electron correlations and antiferromagnetic (AF) order.

The parent Mott insulator itself is relatively well-understood. It is a charge transfer insulator due to the hybridization of the O $2p$ state and Cu $3d$ state [5]. The charge transfer gap (CTG) between the charge transfer band and upper Hubbard band (UHB) has been measured by optical spectroscopy [6,7], resonating inelastic x-ray spectroscopy [8,9], and scanning tunneling microscopy (STM) [10,11]. The electronic structure of the superconducting phase has also been thoroughly investigated by various probes, especially angle-resolved photoemission spectroscopy [12]. The ground state of the superconducting phase is rather conventional, except that the energy gap has a $d$-wave symmetry [13]. The real missing link lies in the crossover regime between the parent compound and the superconducting phase, when dilute holes are dispersed in the AF Mott insulator. The local electronic states of an individual hole dopant and the interactions between them may hold the key for the genesis of Cooper pairing in cuprates.

STM is an ideal technique to explore the few-hole doped cuprates due to its capability of directly visualizing the atomic-scale structural and electronic properties [10,11,14-19]. In this letter, we use STM to image the local electronic states induced by a pair of Ca-vacancy hole dopants in $Ca_2CuO_2Cl_2$ (CCOC) parent cuprate. We find that when the two dopants approach each other, the transfer of spectral weight from UHB to low energy in-gap states creates a broad peak and nearly V-shaped gap around the Fermi level ($E_F$). The peak position shows a sudden drop for distance around 4 $a_0$, where $a_0$ is the lattice constant of the $CuO_2$ plane, and then remains almost constant. The in-gap states exhibit distinct spatial patterns that depend on the configuration of the two dopants relative to the Cu lattice. These results provide important new clues for elucidating the evolution of electronic states in doped Mott insulators.

The STM experiment is performed in an ultrahigh vacuum system with electrochemically etched tungsten tip. The tip is treated and calibrated through strict procedures to ensure stability and reliability [10]. Because the sample is highly insulating at low temperature, all experimental data are acquired at $T$ = 77 K when there are thermally activated carriers. The typical sample bias voltage is $V_b$ = -2.5 V so it is outside the CTG with finite density of state (DOS). The STM topography is taken in the constant current mode with typical tunneling current $I_t$ = 10 pA, and the $dI/dV$ spectra are collected using a standard lock-in technique with modulation frequency $f$ = 447 Hz.

The schematic crystal structure of CCOC is shown in Fig. 1(a). It is an ideal Mott insulator with stoichiometric chemical composition and undistorted $CuO_2$ plane [20]. The crystal can be easily cleaved between two neighboring Cl layers with weak van der Waals bonding. From inductively coupled plasma analysis, as-grown CCOC contains a trace amount of Ca vacancies. Figure 1(b) displays the topography of cleaved CCOC taken with $V_b$ = -2.5 V and $I_t$ = 10 pA. Most of the area has regular square lattice of the surface Cl atoms lying directly above the Cu sites in the $CuO_2$ plane. There are mainly two types of defects. The dark cross-like defects are surface Cl vacancies, while the bright plaquettes are Ca vacancies at the center of four Cl atoms (Fig. 1(d)). Because each Ca atom donates two electrons, a Ca vacancy effectively donates two holes into the $CuO_2$ plane, similar to the role of interstitial oxygen in some more commonly studied cuprates.

We first start from a single dopant. Figure 1(c) shows a series of $dI/dV$ spectra taken around an isolated Ca vacancy with no other defects nearby. At 6 $a_0$ from the vacancy center (black curve), the $dI/dV$ spectrum has a well-defined CTG with $\Delta_{CTG}$ = 2.0 eV, which is the same as that of pristine CCOC [6,10]. At the center of the vacancy (red curve), the main spectral features are a broad peak around $V_b$ = 1.4 V within the CTG, whereas the DOS of the UHB is strongly suppressed. This spectral weight transfer from the high-energy UHB to the low-energy in-gap state is a characteristic feature of doped Mott insulator [15,21]. Further away from the defect center, the height of the in-gap state drops systematically, and the spectral weight of UHB gradually recovers. Figure 1(e) displays the $dI/dV(\mathbf{r}, V_b)$ map taken at 1.4 V, which directly visualizes the spatial distribution of in-gap state. The electronic cloud has an isotropic shape with radius around 4 $a_0$, thus is strongly localized in space.

We then study the electronic structure induced by a pair of Ca vacancies. Figure 2(a) is the topography of two Ca vacancies separated by 8 $a_0$ along the longitudinal direction and 1 $a_0$ along the transverse direction, thus is designated as (8, 1). Figure 2(b) displays the *dI/dV* curves in three representative locations marked by the colored dots in 2(a). Right on Ca vacancies, the *dI/dV* curves show a broad in-gap state like that of isolated Ca vacancy, but the peak position moves down to $V_b$ = 0.7 V. At the midpoint between them (red curve), there is a much weaker in-gap state around 0.7 V, accompanied by smaller spectral weight transfer from the UHB. Figures 2(c) and 2(d) are *dI/dV* maps taken at 0.45 V and 1.4 V, both exhibiting a circular shape with radius ~ 4 $a_0$. Therefore, interaction between the two dopants with (8, 1) configuration pushes the in-gap state towards $E_F$ without modifying the spatial distribution of wavefunctions.

When two Ca vacancies get closer to form a (6, 1) pair, as shown in Fig. 2(e), the spectral features (Fig. 2(f)) are similar to that of the (8, 1) configuration, namely with a broad in-gap state around 0.7 V. But a closer examination of the *dI/dV* spectra and maps reveals some subtle yet important differences. Firstly, the in-gap state at the midpoint (red curve) becomes much more pronounced with height comparable to that on the Ca sites. Secondly, for bias range below 0.45 V, the DOS at the midpoint is larger than that on the Ca sites. These points can be directly visualized by the *dI/dV* maps in Fig. 2(g) and 2(h). Although the 1.4 V map looks similar to that of the (8, 1) configuration, the midpoint of the 0.45 V map becomes much brighter.

When the pair distance is reduced to 4 $a_0$, as shown in Fig. 2(i) for the (4, 0) configuration, more fundamental changes start to appear. As shown in Fig. 2(j), the in-gap peak position moves down significantly to $V_b$ = 0.45 V. Because the DOS at $E_F$ remains zero, an asymmetric V-shaped gap reminiscent of the pseudogap emerges [14,15]. Moreover, the peak height at the midpoint becomes higher than that on the Ca site, indicating strong overlap between the in-gap state wavefunctions from the two dopants. This is clearly visualized by the *dI/dV* map at 0.45 V (Fig. 2(k)), which has an elliptical shape connecting the two Ca vacancies with the brightest point at the middle. More surprisingly, a double-bud structure appears in the *dI/dV* map at 1.4 V (Fig. 2(l)), indicating that at such energy the in-gap state wavefunction has a node along the Ca-Ca direction. This is consistent with the *dI/dV* curves in Fig. 2(j), which reveal that at 2 $a_0$ perpendicular to the Ca-Ca centerline (blue curve) the DOS at 1.4 V is larger than that at the Ca site (green cure) or the midpoint (red curve).

The trend continues for the (3, 0) configuration, as imaged in Fig. 3(a) and schematically illustrated in Fig. 3(c). The in-gap state peaks on the three sites are still around 0.45 V (Fig. 3(b)) and get even more pronounced, making the V-shape gap sharper and more symmetric. The spectrum at the perpendicular direction (green curve) exhibits two well-defined peaks at 0.45 V and 1.4 V. Consistent with these spectral features, the *dI/dV* map at 0.45 V becomes more concentrated at the midpoint, whereas the line node in the 1.4 V map becomes more evident. However, for two Ca vacancies separated by similar distance but with different orientation, totally different spatial patterns are observed. Figure 3(f) shows the topography of two Ca vacancies with a (2, 2) configuration, which is rotated by 45 degree with respect to the Cu-O bond direction (Fig. 3(h)). The low-energy spectra shown in Fig. 3(g) are similar to that of the (3, 0) configuration, exhibiting a nearly symmetric V-shaped gap with the DOS peak around 0.45 V. The in-gap states around 1.4 V, on the other hand, are in sharp contrast to that of the (3, 0) configuration. The spectra on the three representative locations are qualitatively the same, with non-zero DOS extending to the UHB but without the double-peak feature along the perpendicular direction. The $dI/dV$ map in Fig. 3(i) has an elliptical shape for $V_b$ = 0.45 V, and the 1.4 V map in Fig. 3(j) is less elliptical but without an observable node line. The relationship between the node line feature and the pair orientation is also confirmed in supplementary Fig. S1 and Fig. S2.

Figure 4(a) summarizes the electronic structures for Ca-vacancy pairs with different configurations. The general trend is that with decreasing pair distance, the in-gap peak position moves towards lower energy, leading to a narrower and more symmetric V-shaped gap around $E_F$. However, the evolution is not in a continuous manner, and can be roughly divided into two groups. As summarized in Fig. 4(b), the peak position is always around 0.7 V for Ca-Ca distance between 5 $a_0$ and 8 $a_0$. But when the distance is decreased to near 4 $a_0$, the peak position jumps abruptly to about 0.45 V, and remains there for distance down to 2.8 $a_0$ for the (2, 2) configuration. The local hole density for different pair configurations can be estimated by the parallel shift of negative-bias spectrum corresponding to the charge transfer band [22,23], or Zhang-Rice singlet band [24]. As shown in supplementary Fig. S3, the local hole density increases systematically with decreasing pair distance.

Each Ca vacancy studied here represents a hydrogen-like atom in CCOC, and a pair of Ca vacancies can be viewed as the formation of hydrogen molecule in the Mott insulator background. They apparently form the basis for the electronic structure evolution of cuprates [25]. Unfortunately, theoretical study along this direction is rather scarce. Therefore, we will only give a summary of the main findings and list the puzzles awaiting theoretical explanations.

The first finding is the transfer of spectral weight from high energy to low energy, leading to a peak within the CTG. This is characteristic of doped charges in Mott insulator, as has been interpreted theoretically [26-28]. However, although the local hole density increases continuously with decreasing pair distance, the peak position shows a sudden drop near 4 $a_0$.

Interestingly, 4 $a_0$ seems to be a magic length scale in cuprates. It is not only the radius of the electronic cloud for a single Ca vacancy here, but also the periodicity for the well-known checkerboard charge order in various underdoped cuprates [17,18,29-31]. The connection between these different phenomena remains unknown, but they may reflect the same intrinsic length scale associated with doped holes in the parent Mott insulator.

The second finding is the peculiar spatial characters of the in-gap states. The low-energy states around 0.45 V always have an elliptical shape between Ca-Ca pair. The high-energy states around 1.4 V, in contrast, exhibit a node line for the (4, 0) and (3, 0) configurations but not for the (2, 2) configuration with 45-degree rotation. Although the nature of the in-gap states for a pair of dopants is still elusive, there are at least two possible origins for the close relation between its spatial distribution and the underlying $CuO_2$ plane. One is the effect of AF background on the electronic states formed by the hybridization of two hole-dopants, and another is the spatial arrangement of the oxygen orbitals that host the doped holes.

The third finding is the emergence of a V-shaped gap with decreasing pair distance. The overall trend is highly analogous to that in lightly doped $Bi_2Sr_2CuO_{6+\delta}$ (supplementary Fig. S4 and S5) and Na-CCOC [14,15], demonstrating the universality of such behavior. A crucial question is whether this state can be regarded as a precursor of the pseudogap in underdoped cuprate. If that is the case, our results suggest that the pseudogap is the local consequence of doping a Mott insulator, originated from the transfer of spectral weight to low energy and the suppression of DOS at $E_F$. It is fundamentally different from another group of thought that ascribes the pseudogap to the opening of a gap on the Fermi surface by the formation of density wave orders.

In summary, our STM data provide important new clues regarding the local electronic states induced by a pair of hole dopants in parent cuprate. The spectral features and peculiar spatial distributions show systematic evolutions with the configurations of Ca-vacancy pairs. There are also unexpected puzzles that can be used as test grounds for available theories for doped Mott insulators.

**Acknowledgements:** This work was supported by the MOST of China grant No. 2017YFA0302900 and the Basic Science Center Project of NSFC under grant No. 51788104. It is supported in part by the Beijing Advanced Innovation Center for Future Chip (ICFC).

**Figure Captions:**

FIG. 1. (a) Schematic crystal structure of CCOC. The crystal can be easily cleaved between two adjacent Cl layers, as indicated by the grey planes. (b) Topographic image (100 Å × 100 Å) taken at $V_b$ = -2.5 V and $I_t$ = 10 pA. (c) A series of $dI/dV$ spectra taken at locations around a single Ca vacancy, as indicated by the colored dots in topography. (d) Schematic topview of the exposed surface showing the position of the single Ca vacancy (dark cross). (e) The $dI/dV$ map at 1.4 V in the same area as the inset in (c).

FIG. 2. (a), (e), (i) Topographic images around a pair of Ca vacancies with configurations (8, 1), (6, 1), and (4, 0), respectively. (b), (f), (j) The $dI/dV$ spectra taken on three representative locations around the three types of Ca-vacancy pairs as indicated by the colored dots. (c), (g), (k) The $dI/dV$ maps of the three configurations taken at $V_b$ = 0.45 V, and (d), (h), (l) at 1.4 V.

FIG. 3. (a), (f) Topographic images around a pair of Ca vacancies with configurations (3, 0) and (2, 2), respectively. (b), (g) The $dI/dV$ spectra taken on three representative locations around the two types of Ca vacancy pairs as indicated by the colored dots. (c), (h) Schematic structure of the two configurations with respect to the $CuO_2$ lattice and AF order. (d), (i) The $dI/dV$ maps of the two configurations at $V_b$ = 0.45 V, and (e), (j) at 1.4 V.

FIG. 4. (a) The $dI/dV$ spectra taken on the Ca vacancy site for 9 different configurations. (b) Evolution of the in-gap state peak position with the pair distance.

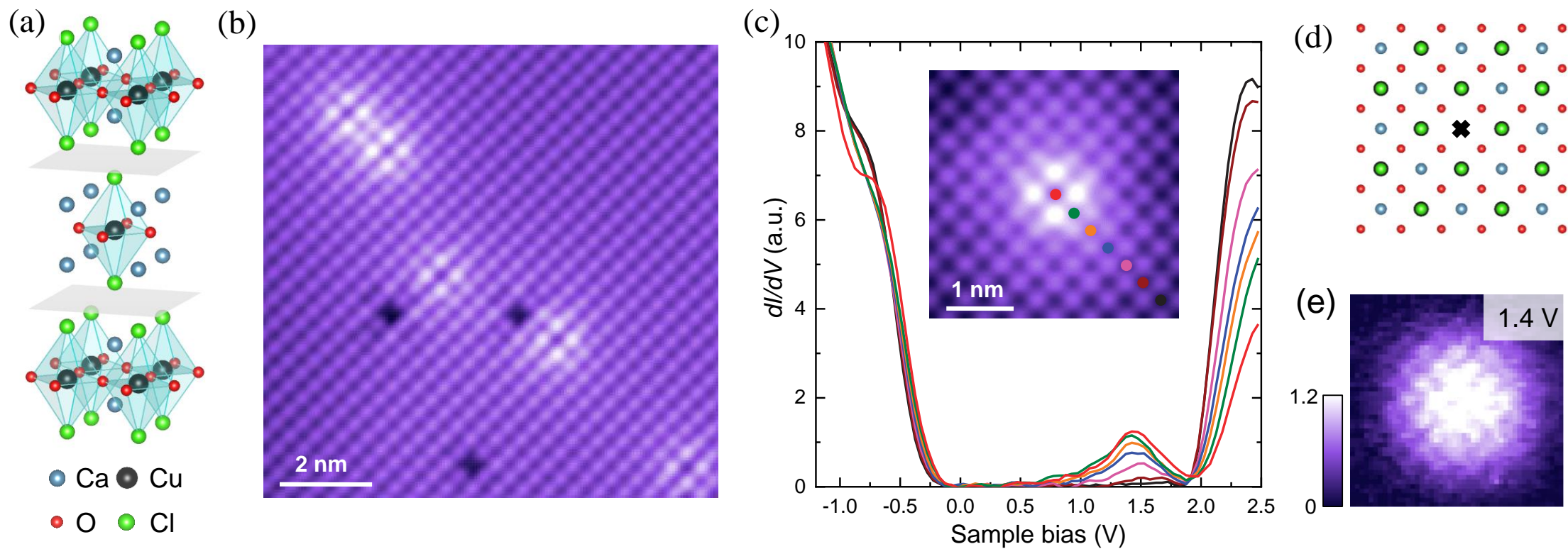

Figure 1

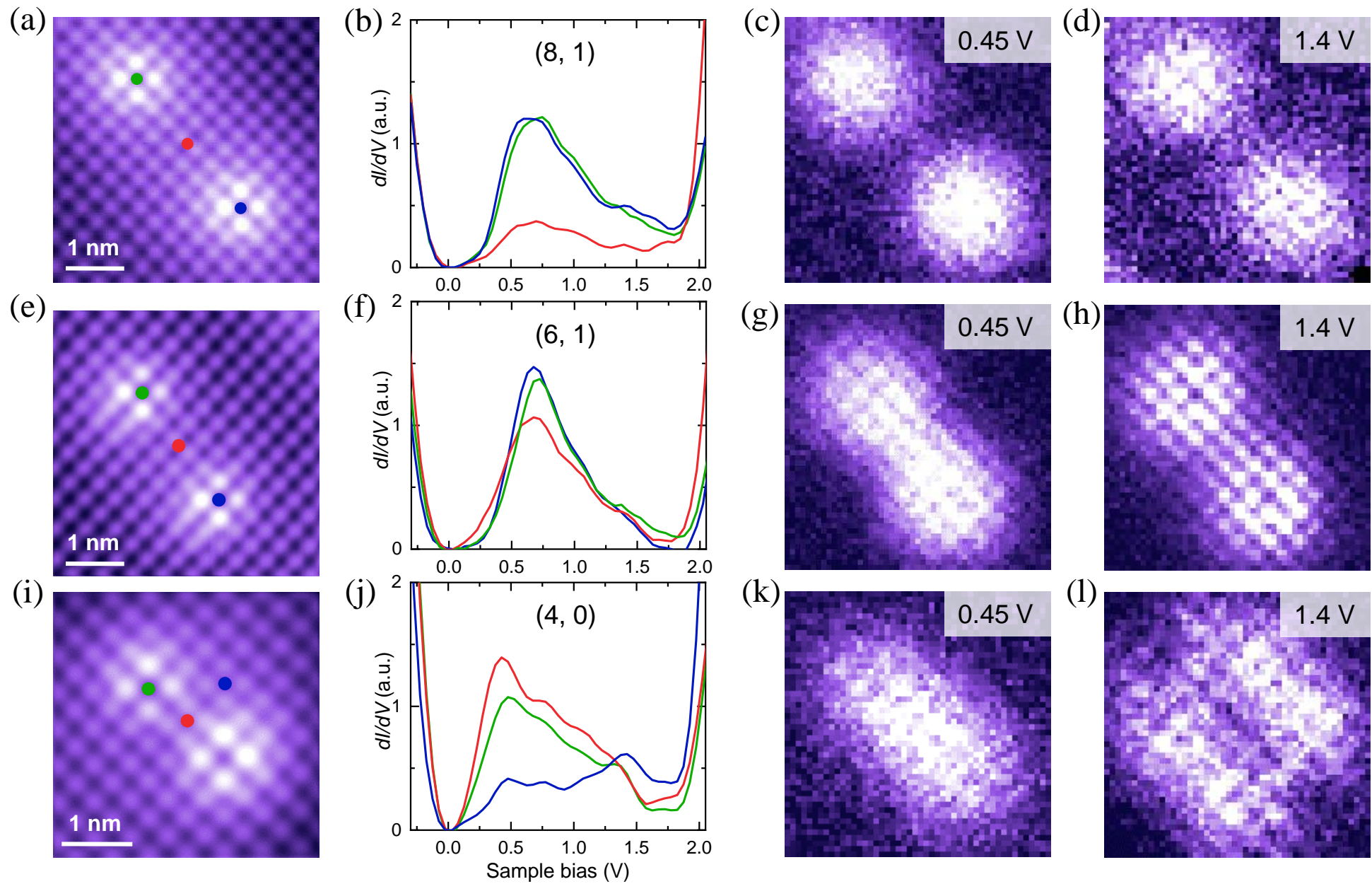

Figure 2

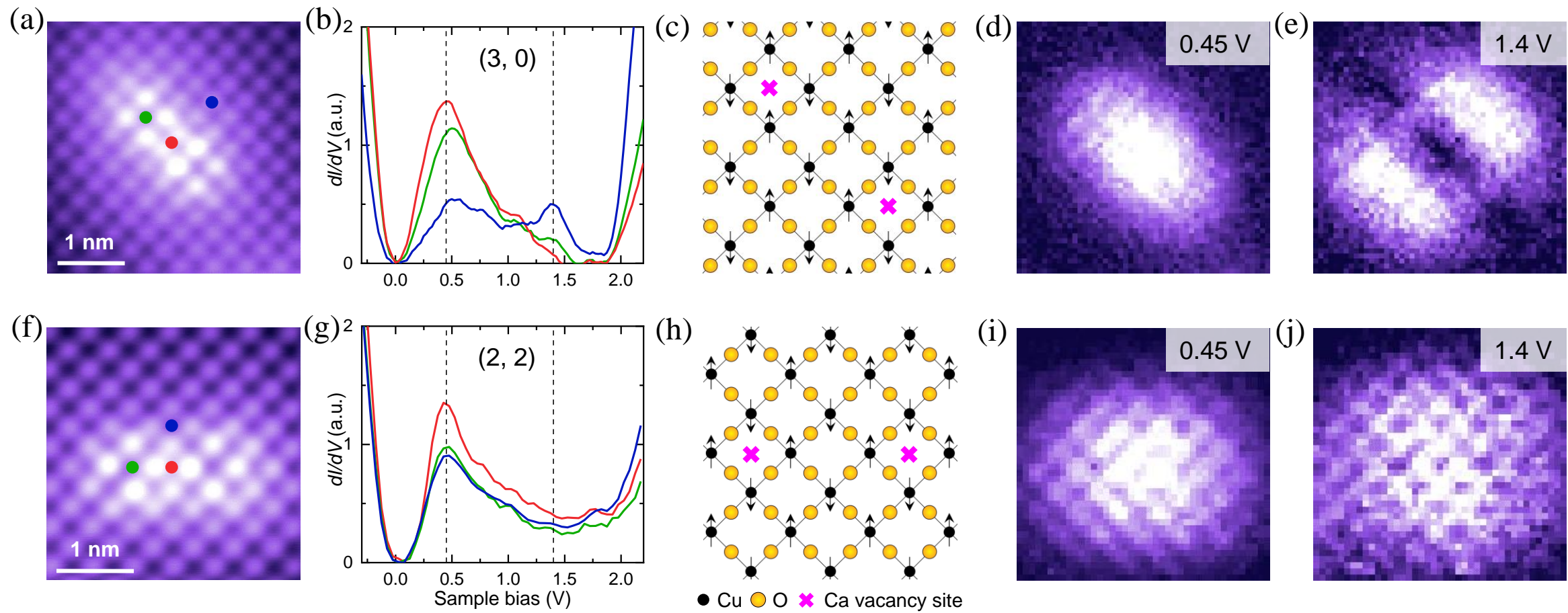

Figure 3

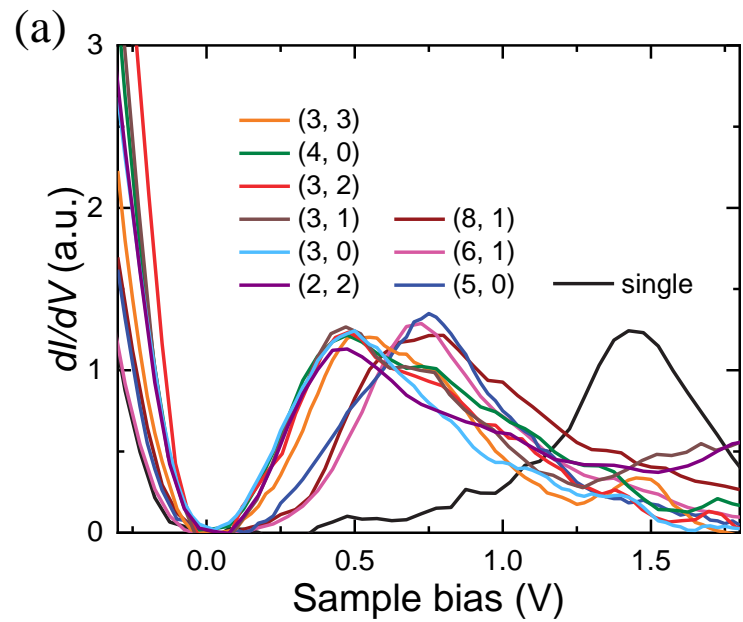 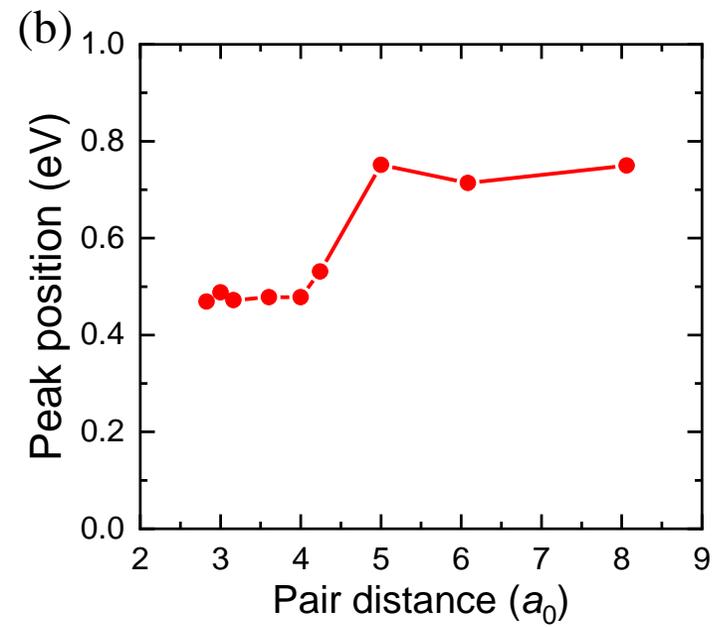

Figure 4